\documentstyle[amssymb,secnumtab,epsfig]{fbssuppl}

\title{Few-Body States in Fermi-Systems and Condensation Phenomena}
\author{P. Schuck$^a$\thanks{{\it E-mail address:} 
schuck@isnhp4.in2p3.fr}, M. Beyer$^b$, G. R\"opke$^b$, W. Schadow$^c$, 
and A. Schnell$^d$}
\institute{$^a$Institut de Science Nucl\'eaires, Universit\'e
Joseph Fourier, CNRS-IN2P3 53, Avenue des Martyrs, F-38026 Grenoble
Cedex, France\\ $^b$Fachbereich Physik, Universit\"at Rostock, D-18051
Rostock, Germany \\
$^c$TRIUMF, 4004 Wesbrook Mall, Vancouver, B.C., V6T 2A3, Canada\\
$^d$Dept. of Physics, University of Washington, 22901, U.S.A.}

\newcommand{\ga}{\alpha}
\newcommand{\gb}{\beta}
\newcommand{\gc}{\gamma}

\begin{document}

\renewcommand{\textfraction}{0.0001}
\renewcommand{\bottomfraction}{0.999}
\renewcommand{\topfraction}{0.9995}

\maketitle
\begin{abstract}
  Residual interactions in many particle systems lead to strong
  correlations. A multitude of spectacular phenomenae in many particle
  systems are connected to correlation effects in such systems, e.g.
  pairing, superconductivity, superfluidity, Bose-Einstein
  condensation etc.  Here we focus on few-body bound states in a
  many-body surrounding.

\end{abstract}

\section{Introduction}

Correlated many particle systems such as, e.g. nuclear matter or
strongly coupled plasmas, have a complicated dynamical behavior.  Only
few areas of the density-temperature phase diagram used to
characterize the state of the system in thermal equilibrium may be
described in the approximation of noninteracting quasiparticles. The
dynamics of the quasiparticle is determined by the mean field of the
other particles and therefore some prominent features like self-energy
corrections and Pauli blocking (Bose enhancement) are sufficiently
taken into account~\cite{fet71}. However due to sizable residual
interactions many interesting and exciting phenomena, such as
clustering, formation of condensates, and phase transitions occur. The
description of the dynamics of such systems is particularly difficult
since the quasiparticle approach reaches its limits.

A large number of these phenomena such as, e.g. two-particle bound
state formation, can already be accounted for by explicitly
introducing two-body correlations into the formalism. This may be
achieved in the frame work of the Green function method~\cite{fet71}
and leads to effective two-body equations that include medium effects
in a consistent way.

\begin{figure}[tb]
\begin{minipage}{0.48\textwidth}
\psfig{figure=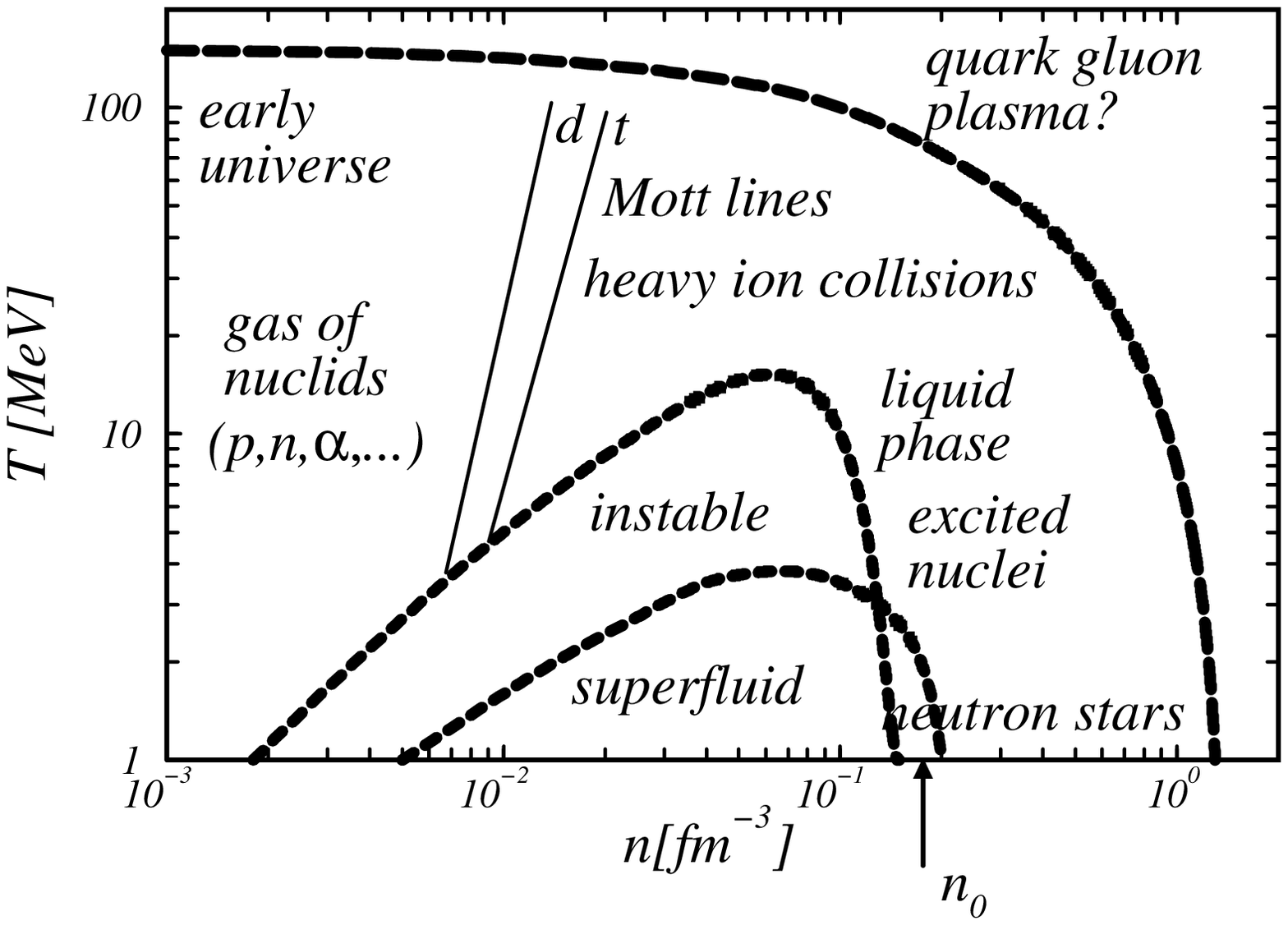,width=\textwidth}
\caption{\label{fig:nt_proc} Density temperature phase diagram of
  nuclear matter. $n_0$ denotes nuclear matter density $n_0=0.17$
  fm$^{-3}$.} 
\end{minipage}
\hfill
\begin{minipage}{0.48\textwidth}
\psfig{figure=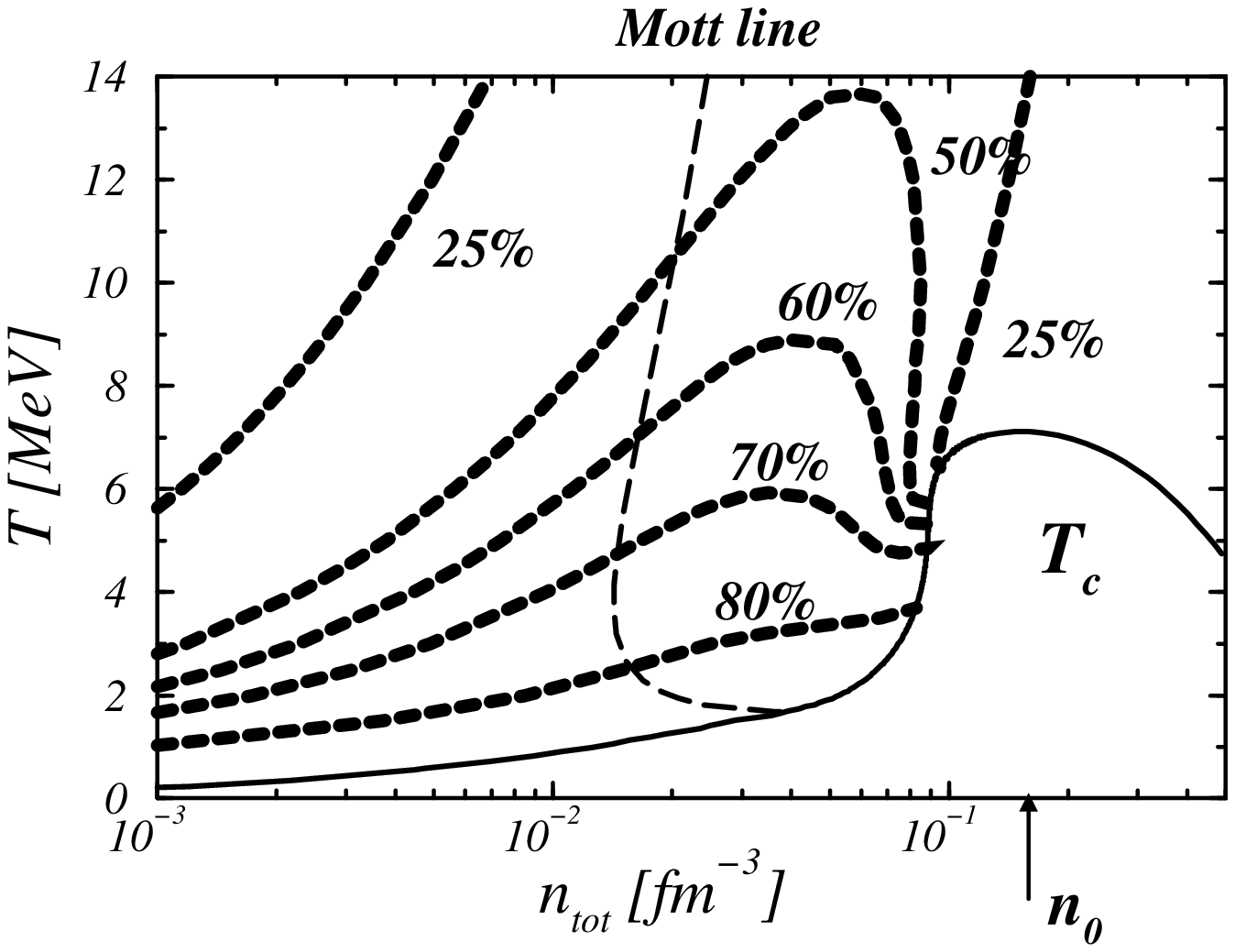,width=\textwidth}
\caption{\label{fig:corr} Nuclear matter phase diagram. The
  the amount of correlated density 
  is given in per cent of the total
  density~\protect\cite{stein}. $T_c$ denotes the critical temperature.}
\end{minipage}
\end{figure}

The density temperature planes of  nuclear matter are shown
in Figs.~\ref{fig:nt_proc} and \ref{fig:corr}. The phase diagram of
nuclear matter turns out to be very rich. In particular the superfluid
phase reflecting strong pairing is relevant for the structure of
neutron stars~\cite{pin92}. At lower densities bound states occur.
This part of the phase diagram may be accessed in the laboratory
through heavy ion collisions at intermediate energies. The description 
of these phenomena in the framework spelled out here constitues a wide 
and exciting area for an application of few-body methods. 

Besides the pairing of nucleons, correlations with more particles are
also important.  Based on the suitably modified AGS
(Alt-Grassberger-Sandhas~\cite{AGS}) formalism the reaction cross
section for break-up ($Nd\rightarrow NNN$), relevant e.g. for the
formation of deuterons in a heavy-ion reaction, has been given
in~\cite{bey96}. As a first application the life time of deuteron
fluctuations in nuclear matter was calculated in~\cite{bey97}.  In
\cite{alpha} we have given an exploratory calculation of the four-body
condensate in strongly coupled fermionic systems.

\section{Theory}
The formalism to derive few-body Green functions within the Dyson
equation approach or cluster mean field approximation at finite
temperatures and densities has been given elsewhere~\cite{duk98}. Here
we give some of the basic results and focus on bound states.

Let the Hamiltonian of the
system be given by
\begin{equation}
H(t) =\sum_{1} H_0(1)a_1^\dagger(t) a_{1}(t)+\;\frac{1}{2}\;
\sum_{12 1'2'}V_2(12,1'2')\;a^\dagger_1(t)  a^\dagger_2(t)  a_{2'}(t)
a_{1'}(t) 
\end{equation}
where $a_1(t)$ etc. denotes the Heisenberg operator of the particle
with quantum numbers $s_1$, $k_1$, etc. for spin, momentum etc. In the
equal time formalism we may introduce creation and annihilation
operators $A_\ga(t)$, where $A_\ga=a_1,a^\dagger_1a_2,a_1a_2a_3$, etc.
for one-particle, hole-particle, three-particle operators. We describe
the propagation of these clusters via the Green's function at
$T=0$. However, generalization to $T\neq 0$ is straight forward using
the Matsubara technique~\cite{fet71}. The definition of the Green's
function reads
\begin{equation} 
{\cal G}_{\alpha\beta}^{t-t'}=-i\langle T A_\ga(t)
A_\gb^\dagger(t')\rangle,
\end{equation}
where averaging is due to the density operator $\rho_0$. For an open
system in thermodynamical equilibrium the extremum condition for the
entropy leads to the following expression for the density operator of
a grand canonical quantum ensemble,
\begin{equation}
\rho_0=\frac{e^{-\gb(H-\mu\,N)}}{{\rm Tr}\{e^{-\gb(H-\mu\,N)}\}}.
\end{equation}
The temperature ($1/\gb=T$) and the chemical potential $\mu$ are 
the corresponding Lagrange parameters. The  Dyson equation is
\begin{eqnarray} 
i\partial_t {\cal G}^{t-t'}_{\ga\gb}
&=&\delta(t-t')  \langle [A_\ga,A_\gb^\dagger]_\pm\rangle
+ \sum_{\gc}\int d\bar t\; { {\cal M}^{t-\bar t}_{\ga\gc}}
\;{\cal G}^{\bar t-t'}_{\gc\gb}.
\end{eqnarray}
We use ${\cal N}_{\ga\gb}= \langle [A_\ga,A_\gb^\dagger]_\pm\rangle$
in the following, where the upper (lower) sign reflect Fermi (Bose)
type operators. The mass matrix is given by~\cite{duk98}
\begin{eqnarray} 
 {\cal M}^{t-\bar t}_{\ga\gc}
&=&\underbrace{\delta(t-\bar t) {\cal M}^{t}_{\ga\gc,0}}_{\rm
  cluster-mean-field} 
 +
 \underbrace{{\cal M}^{t-\bar t}_{\ga\gc,\rm irr}}_{\rm
   retardation}\\
( {\cal M}^{t}{\cal N})_{\ga\gb ,0}&=
&\langle [[A_\ga,H](t),A_\gb^\dagger(t)]\rangle\\
( {\cal M}^{t-t'}{\cal N})_{\ga\gb ,\rm irr}&=
&-i \langle T [A_\ga,H](t),[H,A^\dagger_\gb](t')\rangle_{\rm irr}.
\end{eqnarray}
For $A_\ga=a_1$, i.e. the one-particle case, neglecting the term
${\cal M}_{\rm irr}$ (representing all retarded irreducible diagrams)
leads to the self-energy shift in mean field (e.g. Hartree-Fock)
approximation.  For a given potential this constitutes a
selfconsistent problem to determine $\mu$ and $\gb$, to be solved by
iteration
\begin{equation}
f_1\equiv f(\varepsilon_1) = \frac{1}{e^{\gb(\varepsilon_1 - \mu)}+1}, 
\qquad \varepsilon_1 = \frac{k^2_1}{2m}+\sum_{2}V_2(12)f_2.
\end{equation}
For $T\rightarrow 0$ this function
approaches a step function at the respective Fermi momentum.

\section{Two-body correlations, bound states and Cooper pairing}
Nuclear matter cannot be considered as a system of independent quasi
particles but for a large part it is correlated and, e.g. in the
superfluid phase as demonstrated in Fig.~\ref{fig:corr}~\cite{stein}. 
The basis to treat correlated densities is provided by a
generalization of the Beth-Uhlenbeck approach~\cite{sch90}. The
nuclear density $n=n(\mu,T)$ is given by
\begin{equation}
n=n_{\rm free} + n_{\rm corr},\qquad n_{\rm corr} = 2n_2 + 3n_3 +
\dots,
\end{equation}
where $n_{2,3}$ denotes the two, three-particle correlations, present
as bound/scattering states. In first iteration these correlations may
be treated on the basis of residual interactions between the quasi
particles. The exact two particle equations to be solved are known as
Bethe-Goldstone or Feynman-Galitski equations depending on some
details~\cite{fet71}. In ladder approximation the two-body equation
reads
\begin{equation}
(E - \varepsilon_1 - \varepsilon_2)\,
\Psi_2(E) =   (\bar f_1\bar f_2 - f_1f_2)\, V_2\, \Psi_2(E),
\label{eqn:green2}
\end{equation}
where $\bar f=1-f$. The normalization is given by
\begin{equation}
\sum_{12} \tilde \Psi^{*\nu}_{12}
(f_1f_2-\bar f_1\bar f_2) \tilde\Psi^{\nu'}_{12}=\delta_{\nu\nu'},
\end{equation}
where the dual wave functions $\tilde\Psi^\nu$ are defined by
$\Psi^\nu=(1-f_1-f_2)\tilde \Psi^\nu$. Introducing the two-body
$t$-matrix in a standard fashion both bound and scattering states have
been solved~\cite{sch90}. In first approximation $V_2$ of
Eq.~(\ref{eqn:green2}) corresponds to the bare force. However, at
higher densities the force may be substantially screened.

\begin{figure}[t]
\begin{minipage}{0.48\textwidth}
\psfig{figure=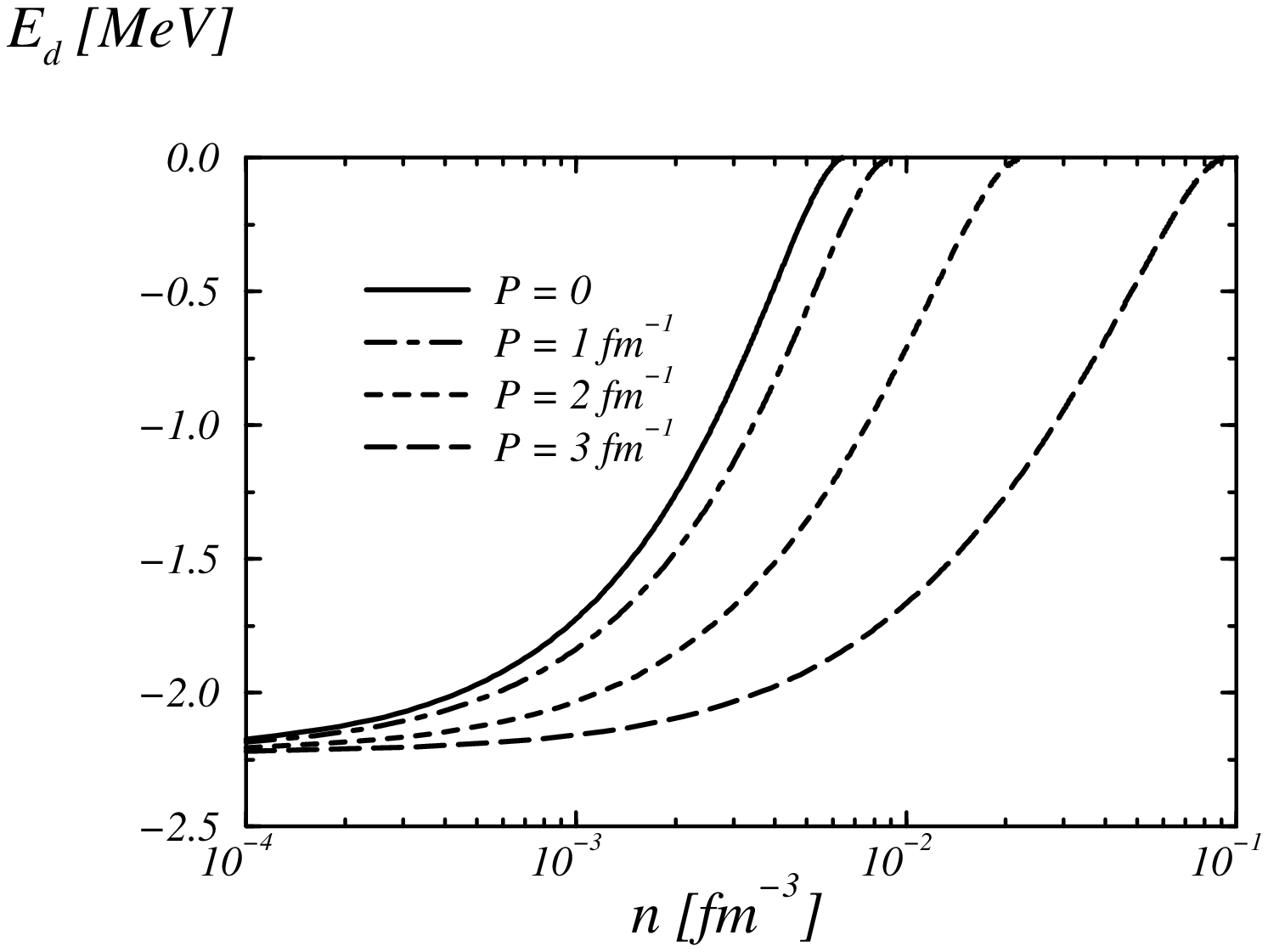,width=\textwidth}
\caption{\label{fig:deut} The deuteron binding energy as a function of 
  the nuclear density for $T=10$ MeV. $P$ denotes the relative momentum 
  between the deuteron and the medium.}
\end{minipage}
\hfill
\begin{minipage}{0.48\textwidth}
\epsfig{figure=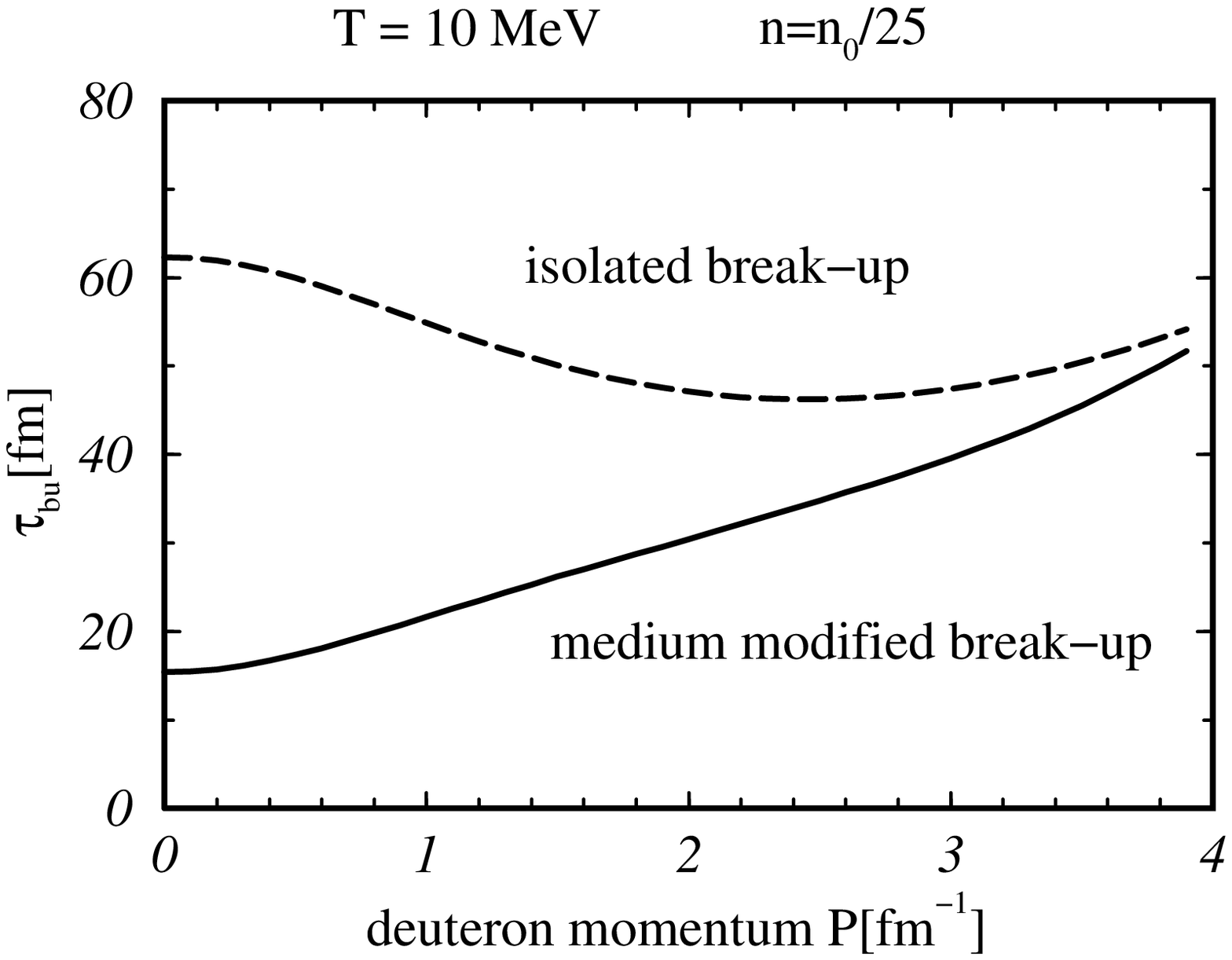,width=\textwidth}
\caption{\label{fig:life} Deuteron life time as a function of the
  deuteron momentum at finite temperature ($T=10$ MeV) and density
  $n=n_0/25$ \protect\cite{kuhrts}.}
\end{minipage}
\end{figure}

As an example for an in-medium bound state the deuteron binding energy
is shown in Fig.~\ref{fig:deut}. The Mott density is defined through
the condition $E_d=0$. Note that the Fermi functions $f_1$ for
particle 1 etc. depend on the relative momentum $P$ between the
deuteron and the medium.

\begin{figure}[thb]
\begin{center}
\epsfig{figure=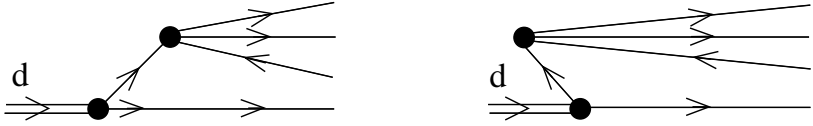,width=0.9\textwidth}
\end{center}
\caption{\label{fig:diagram} Decay of deuteron in nuclear matter
  induced by surrounding nucleons.}
\end{figure}

The deuteron in the medium also has a finite life time. To second
order Born approximation the life time can be extracted from the
square of the graphs of Fig.~\ref{fig:diagram}.
 One sees that the
width $\Gamma_d^\downarrow$ is given by the incoherent decay of the
deuteron into 3p-1h states. At zero temperature one evaluates the
3p-1h density of states
\begin{eqnarray}
g_{\rm 3p-1h}^E&=&{\rm Tr}_1{\rm Tr}_2{\rm Tr}_3{\rm Tr}_4\{
\theta(H_1-\varepsilon_F)
\theta(H_2-\varepsilon_F)
\theta(H_3-\varepsilon_F)
\theta(\varepsilon_F-H_4)\}\nonumber\\
&&\times\delta(E-H_1-H_2-H_3+H_4)\\
&\propto&(E-2\varepsilon_F)^3+\dots
\end{eqnarray}
Since the normalization of the deuteron wave function brings a factor
$(E-2\varepsilon_F)^{-1}$ we have
$\Gamma_d^\downarrow\propto(E-2\varepsilon_F)^2$. At finite
temperatures $T\ll\varepsilon_F$ we get as in Fermi liquid theory for
zero sound a $T^2$ behavior for $\Gamma_d^\downarrow$. In a more
elaborated theory one has to consider rescattering of the incoming
deuteron with the nucleon(s) of the medium. Summing up all the
respective diagrams leads to a generic three-body in-medium scattering
matrix, that has been derived in~\protect\cite{bey96,bey97}. As a
consequence the reaction rate depends strongly on the surrounding
medium.  The numerical result for the life-time of the deuteron is
shown in Fig.~\ref{fig:life}, where we in addition compare the use of
the isolated to the in-medium break-up cross section at $T=10$ MeV and
a nuclear density of $n=0.007$ fm$^{-3}\simeq
n_0/25$~\cite{bey97,kuhrts}.  The medium dependent elementary cross
section leads to shorter fluctuation times. As expected the difference
is vanishing for larger momenta. In turn this effect leads to faster
chemical relaxation times that is e.g. relevant for the notion of
freeze out in heavy ion collisions.

\begin{figure}[tbh]
\psfig{figure=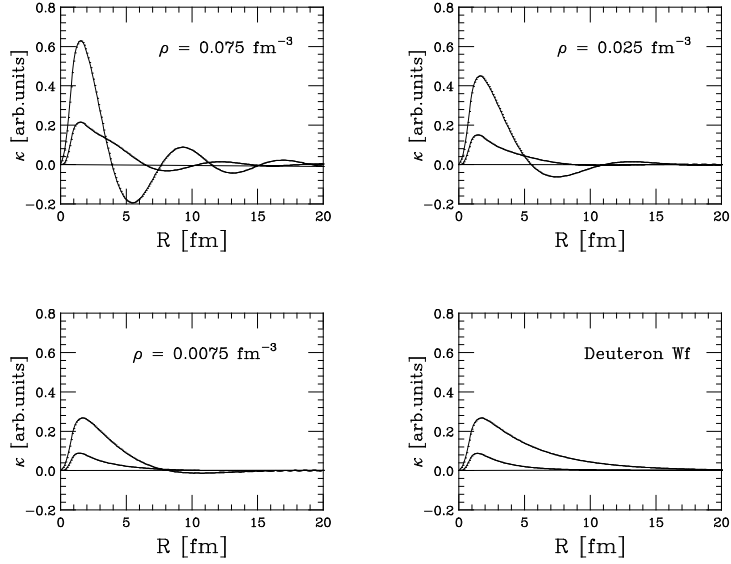,angle=90,width=\textwidth}
\caption{\label{fig:deutfkt} The deuteron (Cooper pair) wave
  function. Large component: s-wave; small component: d-wave. The
  Paris force was used in the gap equation and the single particle
  energies are from Brueckner Hartree-Fock. (From~\protect\cite{bal95}).} 
\end{figure}

The appearance of a sharp pole at $E=2\mu$ in the respective Green
function related to Eq.~(\ref{eqn:green2}) corresponds to the Thouless
criterion for the onset of superfluidity for $np$ Cooper pairs.
Depending on the channels considered, it allows us to determine the
critical temperatures $T_s^c$ or $T_t^c$ for the respective isospin
singlet and triplet channels, as a function of the chemical potential.
This is shown in Fig.~\ref{fig:corr} and later in Fig.~\ref{fig:alpha}
for $T^c_t$. Using the canonical relation $\Delta_{k_F}=1.76 T_c$ one
realizes that the $np$ gap has a maximum value of $\simeq 10$ MeV.
This is more than 3 times larger than the $nn$ gap!  Such high values
may have an influence in expanding nuclear matter as produced in
central heavy ion collisions.

An interesting extension of Eq.~(\ref{eqn:green2}) is to write this
equation down for the case where we have a condensate of pairs, i.e. a
superfluid state. For pairs at rest one has
\begin{equation}
(2\mu-2\varepsilon_p)\Psi_p = (\bar n_p-n_p)
\sum_k v_{pk} \Psi_k,
\label{eqn:deut0}
\end{equation}
where $v_{pk}$ is the two-body interaction in momentum space and the
occupation numbers $n_p$ ($\bar n_p=1-n_p$) are given by
\begin{equation}
n_p=\frac{1}{2}\left[ 1-
\frac{\varepsilon_p-\mu}
{\sqrt{(\varepsilon_p-\mu)^2+\Delta_p^2}}(1-2f_p)\right]
\label{eqn:np}
\end{equation}
with $\Delta_p=\sum_kv_{pk}\Psi_k$ the usual gap function. The
chemical potential $\mu$ and the density are related by
$N/V=2\sum_pn_p$. We see that for a superfluid state
Eq.~(\ref{eqn:green2}) has turned into Eq.~(\ref{eqn:deut0}). Besides
its nonlinearity corresponding to the gap the latter has a similar
structure as the linear Eq.~(\ref{eqn:green2}). One also realizes that
for the zero density limit $N/V\rightarrow 0$ Eq.~(\ref{eqn:deut0})
turns into the Schr\"odinger equation for the deuteron. This is
demonstrated in Fig.~\ref{fig:deutfkt} where we show the Fourier
transform of $\Psi_p$, i.e. the Cooper pair wave function for
decreasing densities (Fig.~\ref{fig:deutfkt} is taken from
~\cite{bal95}). One clearly sees that the $np$ Cooper pair wave
function approaches the free deuteron wave function in the zero
density limit.

\section{Three-body reactions and bound states}
A first three-body equation for quasiparticles embedded in a medium at
zero temperature has been given in~\cite{sch73}. Here we present
the equation for the wave function
\begin{equation}
(E - \varepsilon_1 - \varepsilon_2- \varepsilon_3)
\Psi_3(E)= [(\bar f_1\bar f_2 - f_1f_2) V_2(12) + {\rm perm.}]
\;\Psi_3(E)
\label{eqn:green3}
\end{equation}
together with the normalization 
\begin{equation}
\sum_{123} \tilde \Psi^{*\nu}_{123}
(f_1f_2f_3+\bar f_1\bar f_2\bar f_3) 
\tilde\Psi^{\nu'}_{123}=\delta_{\nu\nu'},
\end{equation}
where the dual wave functions are defined by  by 
$\Psi^\nu_{123}=(f_1f_2f_3+\bar f_1\bar f_2\bar f_3) 
\tilde\Psi^\nu_{123}$.

\begin{figure}
\begin{minipage}{0.48\textwidth}
  \leavevmode
\centering
  \psfig{figure=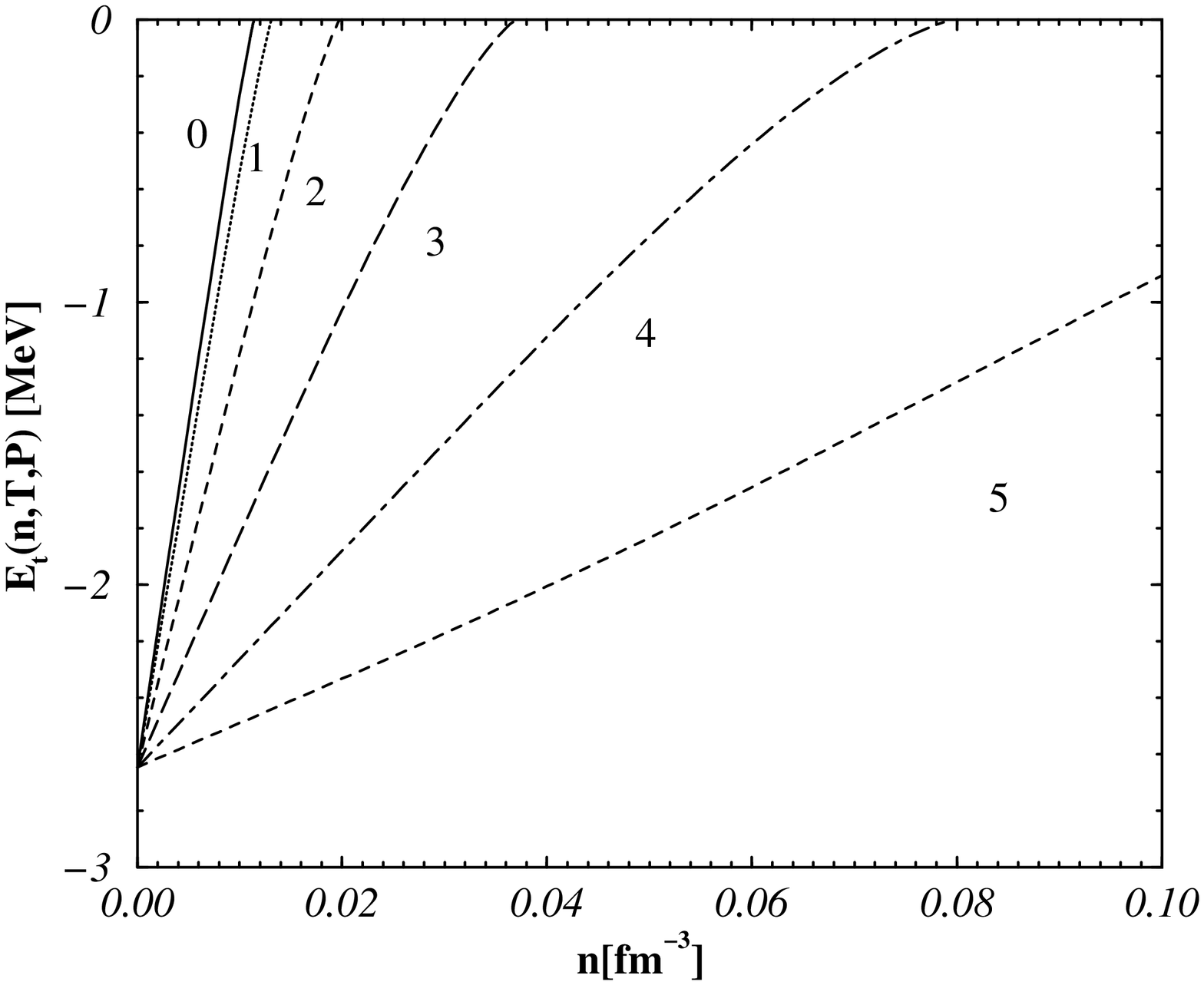,width=0.7\textwidth,angle=270}
\vspace*{1ex}
\caption{\label{fig:triP} Triton binding energy per nucleon as a function of
  nuclear density at $T=10$ MeV, and total momentum relative to the
  medium. From left to right:
  $P=0,1,2,3,4,5$ fm$^{-1}$ \protect\cite{schadow}.}
\end{minipage}
\hfill
\begin{minipage}{0.48\textwidth}
  \leavevmode
\centering
 \psfig{figure=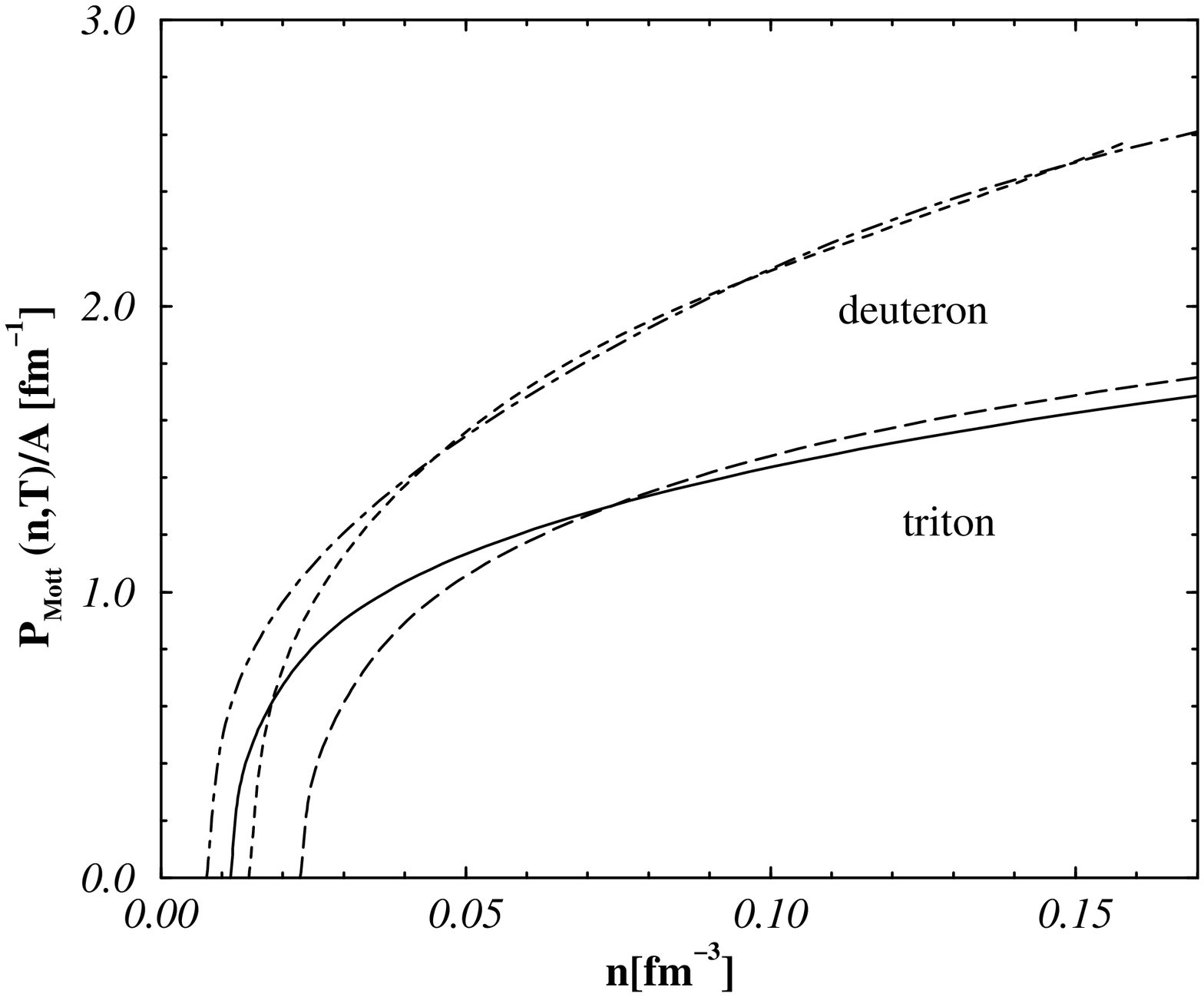,width=0.7\textwidth,angle=270}
\vspace*{1ex}
\caption{\label{fig:mottP} Mott momentum per nucleon $P_{\rm Mott}/A$
  for triton - solid ($T=10$ MeV) and dashed line ($T=20$ MeV) and
  deuteron - dashed dotted ($T=10$ MeV) and short dashed line ($T=20$
  MeV) as a function of nuclear density \protect\cite{schadow}.}
\end{minipage}
\end{figure}

An extension to finite temperatures uses the Alt-Grassberger-Sandhas
(AGS) formalism to treat scattering phenomena and has been given
in~\cite{bey96,bey97,kuhrts}. For a numerical solution of the
scattering problem we use a separable ansatz for the strong
nucleon-nucleon potential~\cite{bey96}. In~\cite{bey96} we have
calculated the break up cross section $Nd\rightarrow NNN$ and found
considerable dependence of the deuteron fluctuation time on the proper
treatment of the medium dependence~\cite{bey97}, see also
Fig.~\ref{fig:life}. The influence of this refined treatment on the
relaxation times is treated in~\cite{kuhrts}. The solution for the
medium modified three-body bound state using the Yamaguchi and the
Paris potential for comparison is provided in~\cite{schadow}.
Fig.~\ref{fig:triP} shows the triton binding energy depending on
different c.m. momenta as a function of the nuclear density and
Fig.~\ref{fig:mottP} shows the Mott momentum. Below the respective
lines no deuterons/triton exist as bound states.

A very interesting situation can also happen when a sharp pole at
$E=3\mu$ appears in the in-medium three-body equation
(\ref{eqn:green3}). Then the original Fermi gas becomes
unstable with respect to the new Fermi gas composed out of
three-body clusters. For a very low density nuclear matter this could 
be a Fermi gas of $^3$He or tritons. Though this case may be somewhat
academic because of the presence of other type of clusters,
e.g. $\alpha$-particles, one may
think of another situation where this consideration may become very
relevant; suppose one cools down a quark gluon plasma to the point
that nucleons (and other hadrons) appear. Then the quark Fermi gas
transforms into a nucleon Fermi gas. A relativistic equation describing this
transition should have a similar structure as
Eq.~(\ref{eqn:green3}). From such an equation the critical temperature 
for the transition from quark gas to nucleon gas could be studied.

\section{Four-nucleon correlations and quartetting}

The four-nucleon correlation is believed to play a significant role
for lower densities and temperatures because of the strong binding
energy of the $\alpha$-particle. Exploratory calculations using a
simple variational ansatz for the $^4$He wave function predict an
$\alpha$ condensate/quartetting on top of the deuteron
condensate/triplet pairing that leads to superfluidity~\cite{alpha}.

Bose systems behave quite differently with respect to the occurrence
of bound states.  The Bose functions enhance the effective residual
interaction that might lead to an ``opposite Mott effect'', i.e.
existence of bound states and also pairing (condensate) even if no
bound state exist for the isolated case. An example is provided by a
pion gas, were a pionic condensate may occur~\cite{alm97} that are
discussed, e.g. in the context of neutron stars~\cite{pin92}.

\begin{figure}
\begin{minipage}{0.48\textwidth}
  \leavevmode
\centering
\psfig{figure=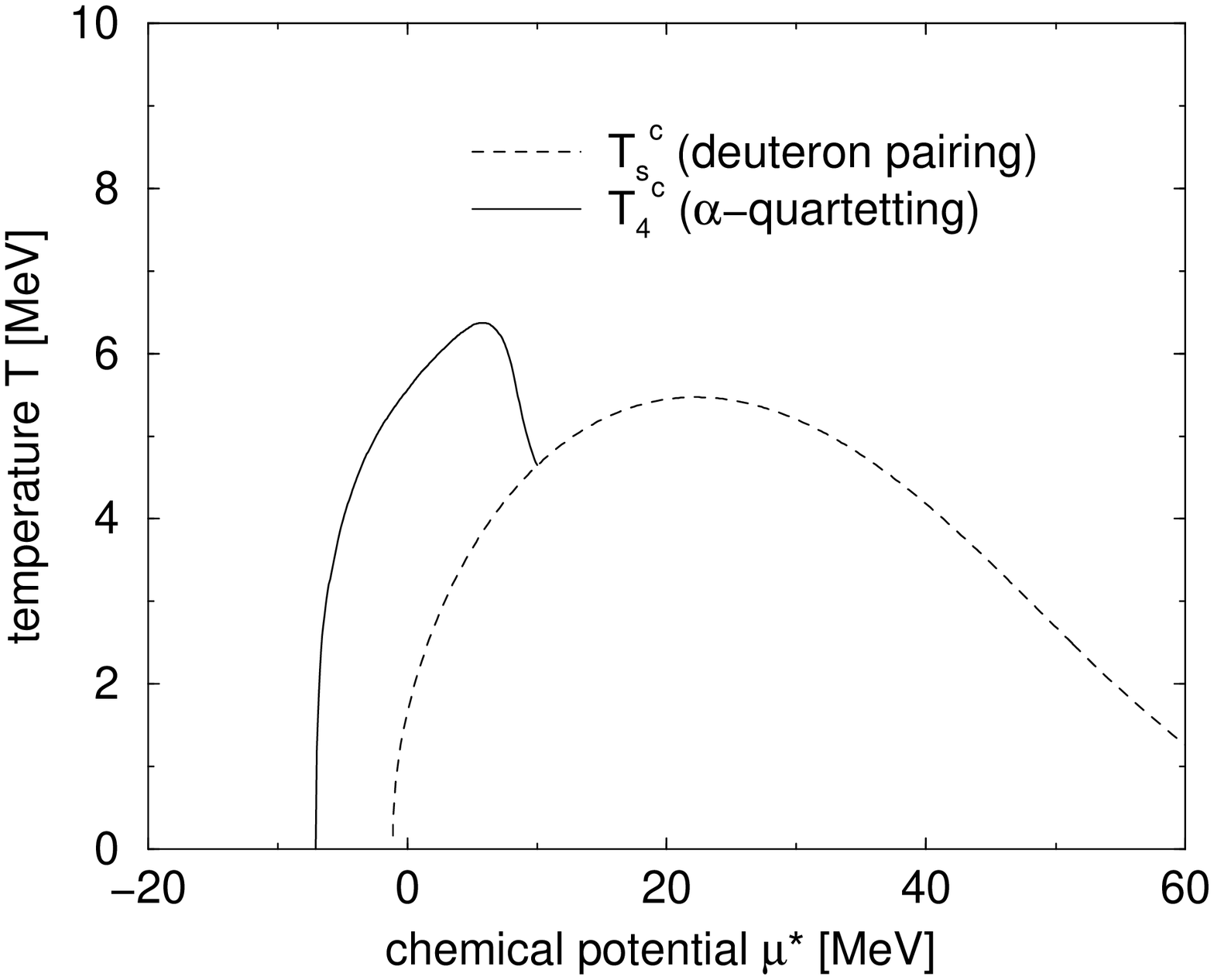,width=\textwidth}
\end{minipage}
\hfill
\begin{minipage}{0.48\textwidth}
  \leavevmode
\centering
\psfig{figure=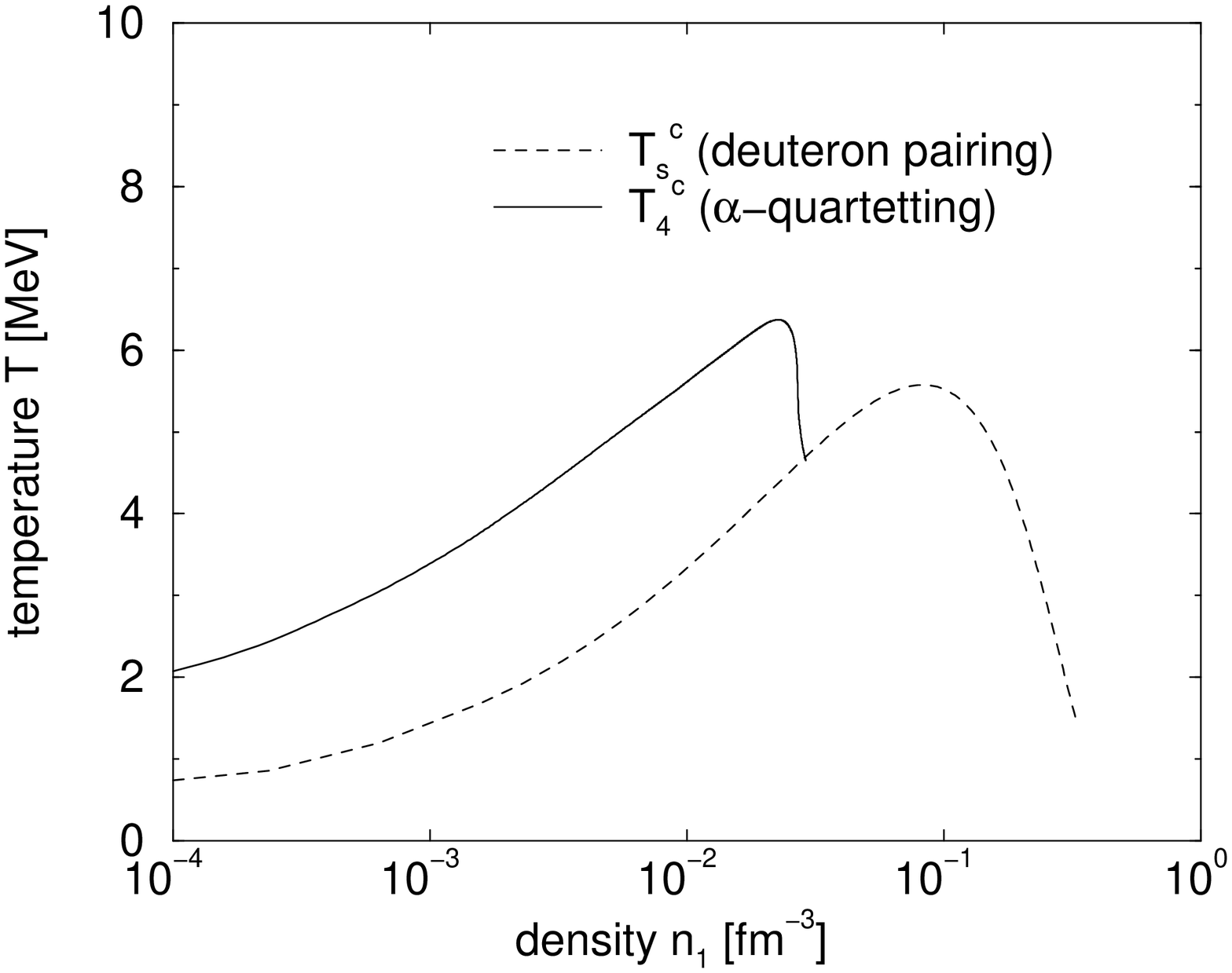,width=\textwidth}
\end{minipage}
\caption{Critical temperatures for the onset of quantum condensation in
symmetric nuclear matter, model calculation. The critical temperature of the
onset of two-particle pairing $T_t^c$ is compared with $T_4^c$ for the onset
of a four-particle condensate, as a function of the chemical potential 
(left) and as a function of the
uncorrelated density (right) \protect\cite{alpha}.}
\label{fig:alpha} 
\end{figure}

 In the framework utilized 
here the four body Schr\"odinger equation for the bound state problem reads
\begin{equation}
(E - \varepsilon_1 - \varepsilon_2- \varepsilon_3- \varepsilon_4)
\Psi^\nu_4(E)= [(\bar f_1\bar f_2 - f_1f_2) V_2(12) + {\rm perm.}]
\;\Psi^\nu_4(E)
\label{eqn:green4}
\end{equation}
with the normalization~\cite{dan94}
\begin{equation}
\sum_{1234} \tilde \Psi^{*\nu}_{1234}
(f_1f_2f_3f_4-\bar f_1\bar f_2\bar f_3\bar f_4) 
\tilde\Psi^{\nu'}_{1234}=\delta_{\nu\nu'},
\end{equation}
and the dual wave functions
$\Psi^\nu_{1234}=(f_1f_2f_3f_4-\bar f_1\bar f_2\bar f_3\bar f_4) 
\tilde\Psi^\nu_{1234}$.

The solution for $\Psi_4(4\mu)$ provides the critical temperature
$T_4^c$ for $\alpha$ condensates as a function of the chemical
potential shown in Fig.~\ref{fig:alpha} (left), or of the density as
shown in Fig.~\ref{fig:alpha} (right).  Using a variational ansatz to
model the four-body wave function the transition to quartetting
outreaches the transition to isospin singlet pairing if the density is
smaller than $0.03~{\rm fm}^{-3}$, see Fig.~\ref{fig:alpha}.  This may
again be relevant in expanding nuclear matter or even in the far tail
of the density of a nucleus.

One may suspect that the asymmetry of nuclear matter hinders the
formation of $np$ pairing or of quartetting. This is because extra
neutrons Pauli-block the formation of Cooper pairs. However, once the
density is so low that the corresponding chemical potentials turn
negative, then the blocking effect, i.e. the Pauli principle is less
effective and formation of deuterons and $\alpha$ particles should not 
be much unfavored in asymmetric matter with respect to the symmetric case.

\section{Conclusion}

We have studied the formation of clusters and their condensation in
Fermi systems using the Dyson equation approach to many-body Green's
functions. They only depend on the c.m. energy of the cluster and
therefore the resulting equations are of the Schr\"odinger type. To
lowest order the in-medium $n$-body equations are -- compared to the
isolated case -- only modified by RPA phase space factors on the
two-body matrix elements (leading to self-energy corrections and
modified interactions). Such equations have early been written down by
the Japanese school~\cite{marumori}. We here investigate them with
respect to the Mott effect and Bose-Einstein condensation in
homogeneous matter. It is found that deuteron Cooper pairs survive
longer than $\alpha$-particle quartets at high densities but that, not
unexpectedly, $\alpha$-particles win in the low density limit. We give
the Mott density for deuterons and tritons and discuss the Pauli
principle aspect of the phase transition of a quark Fermi gas to a
nucleon Fermi gas. Life time effects of the clusters are also
considered. To second order Born approximation they behave like
$\Gamma_n^\downarrow = a\;(E-n\varepsilon_F)^2 +bT^2$, where $n$ is the
nucleon number in the cluster. Therefore $n$-body clusters at
$E_n=n\varepsilon_F$ and $T=0$ have infinite life time.

Although we have taken examples from nuclear physics, the exact
treatment of few-body systems embedded in the medium is not
restricted to nuclear physics. Further potential applications are
possible, e.g. in the field of semiconductors to treat the formation
of excitons or the trion bound state in the dense low dimensional
electron plasmas. For recent experiments see e.g.~\cite{trion}. For
highly ionized dense plasmas the impulse approximation may fail for
the three-particle break-up cross section needed to calculate the
ionization and recombination rates.

\begin{acknowledge}
  It is a pleasure for us to thank M. Baldo, C. Kuhrts, U. Lombardo,
  H. J. Schulze for collaborations and discussions.
\end{acknowledge}

\SaveFinalPage
\end{document}